# Dipole-dipole interactions in protein-protein complexes: a quantum mechanical study of the ubiquitin-Dsk2 complex

Fabio Pichierri[a,b*]

[a]Department of Applied Chemistry, Graduate School of Engineering, Tohoku University,

Aoba-yama 6-6-07, Sendai 980-8579, Japan

[b]Quantum Proteomics Initiative (QPI)

[v.1, August 14, 2013]

Abstract.

Quantum mechanical calculations are performed on the proteins that constitute the ubiquitin-Dsk2 complex whose atomic structure has been experimentally determined by NMR spectroscopy (PDB id 1WR1). The results indicate that the dipole moment vectors of the two proteins are aligned in a head-to-tail orientation while forming and angle of ~130°. Hence, attractive dipole-dipole interactions not only stabilize the protein-protein complex but they are likely to favor the correct orientation of the proteins during the formation of the complex.

[*] Corresponding author. Tel. & Fax: +81-22-795-4132

*E-mail address:* fabio@che.tohoku.ac.jp (F. Pichierri)

Protein-protein (PP) interactions (interactomics) are ubiquitous in biology. For instance, the formation of antigene-antibody complexes is regulated by PP interactions as well as the transmission of signals within and among cells (Jones and Thornton, 1996; Keskin et al., 2008; Nussinov and Schreiber, 2009; Perkins et al., 2010). Electrostatic effects are thought to play an important role in both providing the correct orientation of the proteins as well as in stabilizing the resulting PP complexes (Zhang et al., 2011). In this regard, PP interfaces do often display complementary electrostatic characteristics and, in addition, localized interactions such as H-bonds and salt-bridges are observed to occur among the polar and charged amino acid residues that populate the PP interfaces (Lo Conte et al., 1999).

Although the analysis of the electrostatic potential of PP interfaces is often carried out in the study of PP complexes (Honig and Nicholls, 1995), the contribution from charged residues that are located far from the interfacial region is not properly accounted for by such analyses. This contribution, however, can be included quantitatively when one considers the interaction between the electric dipole moments of each protein in the complex. Protein dipoles can be experimentally measured by electrooptical measurements (Antosiewicz and Porschke, 1989) or with the dielectric constant technique (Takashima, 1996) as well as calculated using different approaches (Pichierri, 2003; Felder et al., 2007). From a purely physical viewpoint, a pair of electric dipole vectors can interact with each other via a parallel or anti-parallel head-to-tail orientation, as schematically shown in Figure 1(a,b). In the case of interaction between protein dipoles, however, both the non-spherical shape of the biomolecules and the formation of localized chemical interactions (H-bonding, etc.) will result in the alignment shown in Figure 1(c) where the dipole vectors form an angle 0° < α < 180°.

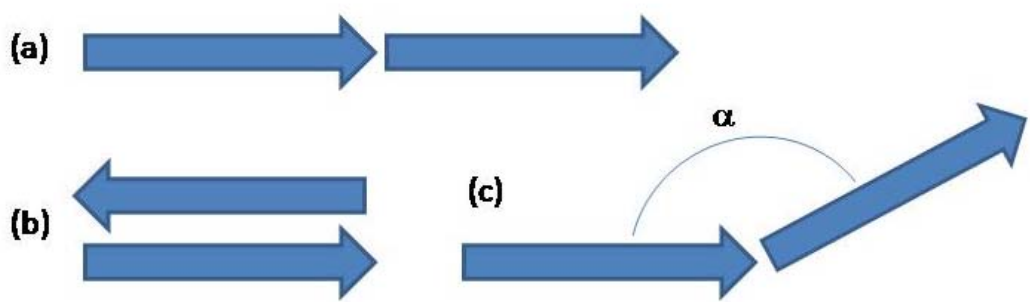

**Figure 1.** Head-to-tail alignment of electric dipoles: (a) parallel and, (b) anti-parallel orientations. (c) Intermediate orientation of dipoles forming and angle 0° < α < 180°. Note that in this study the physical convention is employed for the dipole moment vector according to which the head of the arrow bears a positive charge and the base of the arrow a negative charge.

Here we present the results of quantum mechanical calculations with the MOPAC2009 software package (Stewart, 1996; Stewart 2007; Stewart, 2008; Stewart 2009) aimed at determining the electric dipole moments of ubiquitin (Ubq) and Dsk2 in the Ubq-Dsk2 complex. The atomic coordinates of the NMR structure of the complex (Ohno et al., 2005) which are deposited into the Protein Data Bank (PDB id 1WR1, model 1) were employed. Two independent self-consistent field (SCF) calculations were performed, one on Ubq (charge=0) and another on Dsk2 (charge=−4). The effect of the solvent (water) environment was taken into account with the aid of Conductor-like Screening Model, COSMO (Klamt and Schüürmann, 1993), and by setting the dielectric constant ε at the value 78.4 recommended for water.

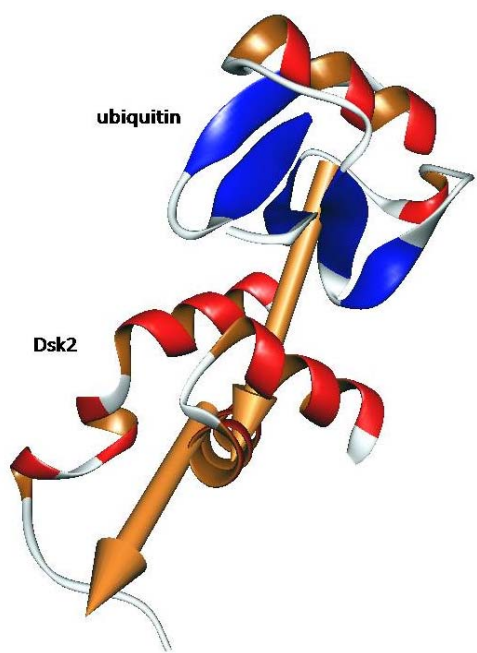

**Figure 2.** Dipole-dipole interaction in the ubiquitin-Dsk2 complex (PDB id 1WR1).

The computed dipole moments of Ubq and Dsk2 have magnitudes of 269 Debye and 204 Debye, respectively. Figure 2 shows the relative orientation of the dipole moment vectors of the two proteins in the Ubq-Dsk2 complex. The two dipole vectors, each located on the centre of mass of the corresponding protein, assume a head-to-tail orientation with the positive arrow's tip of Ubq pointing toward the negative arrow's base of Dsk2 (the angle α between the dipole vectors is ~130º). Hence, this result indicates that the dipoles produce an attractive electrostatic potential which stabilizes the PP complex. Further, this result suggests the possibility that the electrostatic interaction between the dipoles helps the two proteins to achieve the correct orientation that yields the PP complex. An example of dipole-induced orientation effect was recently discussed by the author for the complex between the KcsA potassium channel and the charybdotoxin peptide blocker (Pichierri, 2011). A number of protein-protein complexes which are relevant to biology are currently under investigation and the results will be presented in due course.

### Acknowledgments

I thank Dr. J.J.P. Stewart (Stewart Computational Chemistry, Colorado Springs) for the continuous developments of the MOPAC software package. Dr. Christian Griesinger (Max Planck Institute for Biophysical Chemistry, Göttingen) is acknowledged for suggesting the calculation of the dipole-dipole interaction in the Ubq-Dsk2 complex. This work is supported by the Department of Applied Chemistry of the Graduate School of Engineering, Tohoku University.